**The Role of High Energy Photoelectrons on the Dissociation of Molecular Nitrogen in Earth's Ionosphere**

Srimoyee Samaddar[1], Scott. M. Bailey [1], Justin Yonker[3], Karthik Venkataramani[1,2], Brentha Thurairajah[1]

[1] Center for Space Science and Engineering and Bradley Department of Electrical and Computer Engineering, Virginia Tech, Blacksburg, Virginia, USA.

[2] now at Johns Hopkins University, Pasadena, CA

[3] Johns Hopkins University/Applied Physics Lab in Laurel, MD

Corresponding Author: Srimoyee Samaddar (srimoy1@vt.edu)

**Key Points**

1. Electron impact ionization and excitation cross-sections of major species $N_2$, $O_2$ and O are updated by fitting procedures

2. Updated cross-sections are used in a global average thermosphere/ionosphere model to simulate nitric oxide (NO) production

3. Simulations indicate an increase in NO concentrations by 20% at altitudes below the peak NO

**Abstract**

Soft x-ray radiation from the sun is responsible for the production of high energy photoelectrons in the D and E regions of the ionosphere, where they deposit most of their ionization energy. The photoelectrons created by this process are the main drivers for dissociation of Nitrogen molecule ($N_2$) below 200 km. The dissociation of $N_2$ is one of main mechanisms of the production of Nitric Oxide (NO), an important minor constituent at these altitudes. In order to estimate the dissociation rate of $N_2$ we need its dissociation cross-sections. The dissociation cross-sections for $N_2$ by photoelectrons are primarily estimated from the cross-sections of its excitation states using predissociation factors and dissociative ionization channels. The lack of cross-sections data, particularly at high electron energies and of higher excited states of $N_2$ and $N_2^+$, introduces uncertainty in the dissociation rate calculation, which subsequently leads to uncertainties in the NO production rate from this source. In this work, we have fitted updated electron impact cross-sections data and by applying predissociation factors obtained, updated dissociation rates of $N_2$ due to high energy photoelectrons. The new dissociation rates of $N_2$ are compared to the dissociation rates obtained from Solomon and Qian [2005]. The new dissociation cross-sections and rates are estimated to be about 30% lower than the Solomon and Qian [2005] model. Simulations using a parameterized version of the updated dissociation rates in the Atmospheric Chemistry and Energetics (ACE1D) model leads to a 20% increase in NO density at the altitudes below 100 km is observed.

**Plain Language Summary**

Radiation from the sun in the x-ray and EUV regions of the electromagnetic spectrum is responsible for photoionization of the major neutral species, $N_2$, $O_2$ and O, in the Earth's atmosphere. This process leads to the production of ions and energetic electrons called photoelectrons. In this study, we use a very strong solar flare event to study the process of dissociation of $N_2$ by these photoelectrons and the subsequent production of Nitric Oxide (NO), one of the most important minor constituents. For this study, we focus on the altitude range below 200 km of the ionosphere since this is the region where soft x-rays deposit most of their energies and where the processes that occur as a result of this deposition are most pronounced. In order to have a better understanding of these high energy processes, we have used updated electron impact excitation and ionization cross-sections and predissociation factors.

# 1. Introduction

The soft x-ray (below 30 nm) (XUV) and Extreme Ultraviolet (EUV) (30-120 nm) regions are the most variable and dynamic regions of the solar irradiance spectrum. These regions have the ability to ionize neutral species, primarily $N_2$, $O_2$ and O, in the Earth's D and E layers of the ionosphere. The ionization process also creates energetic photoelectrons that can have enough energy to cause further ionization, excitation and dissociation of neutrals and ions in the atmosphere. Siskind et al. [2022] showed that photoelectrons, created by the soft x-rays, are the main drivers for ionization at lower altitudes, i.e., below 90 km. The effects of these photoelectrons are especially enhanced during solar flares. To study the photoionization and photoelectron ionization during solar flares Siskind et al. [2022] developed the NRLFLARE spectral model with very fine resolution of spectral irradiance in the soft x-ray region. The fine resolution is formed by resolving the altitude of unit optical depth for different photon wavelength bins. This consequently resolves the altitudes of photoelectron ionization at this low altitude. This model improves upon the earlier EUVAC spectrum (Richards et al. [1994]) as described in Solomon and Qian [2005] (hereon referred to as SQ'05) which used three wide bins between 0.05 - 1.8 nm. This coarse resolution cannot be used to represent the solar flare variability, where the photon energy changes as much as eight orders of magnitudes and the flux by almost two orders of magnitudes.

The photoelectrons created at altitudes below 90 km have energies in the range of 10 keVs. The minimum energy to dissociate $N_2$ is 10 eV. Thus, these high energy photoelectrons can dissociate the very stable $N_2$ molecule. They also have the ability to produce higher excited states of $N_2$, which primarily pre-dissociate to give N fragments in ground and excited states. The process of photoelectron dissociation mainly occurs either due to electron impact dissociative ionization as shown in Equations 1a and 1b or dissociative excitation as in Equations 2a and 2b.

$$N_2 + e^* \rightarrow N_2^+ + e^* + e \quad (1a)$$

$$N_2^+ + \rightarrow N^+ + N^* \quad (1b)$$

$$N_2 + e^* \rightarrow N_2^* + e^* \quad (2a)$$

$$N_2^* \rightarrow N + N^* \quad (2b)$$

The excited state N atoms produced due to the dissociation of $N_2$ is an important source of Nitric Oxide (NO) in the atmosphere (Bailey et al., 2002; Barth et al., 1992), as shown in Equation 3.

$$N^* + O_2 \rightarrow NO + O \quad (3)$$

The N atoms are usually in three states: the ground state $N(^4S)$ and the two excited states, $N(^2D)$ and $N(^2P)$. The reaction of $O_2$ with $N(^2D)$ is the primary mechanism of production of NO in the lower altitude below 100 km, while the reaction of $O_2$ with $N(^4S)$ becomes important at higher altitudes (Bailey et al. [2002]). However, $N(^4S)$ is also a dominant loss mechanism for NO at the high altitude F region. The contribution of $N(^2P)$ is about two orders of magnitude less than $N(^2D)$ above peak (106 km) NO production and about an order of magnitude less below 100 km. Other NO production and loss mechanisms have been explained in detail by Barth et al. [1992]. It is well established (Barth et al.,1988; Siskind et al., 1990; Bailey et al., 2002) that there is a direct correlation between soft x-rays flux and production of NO in the E region of the atmosphere. With the improved solar irradiance data obtained from Student Nitric Oxide Explorer (SNOE) satellite (Bailey et al., 1999; 2001), Bailey et al. [2002] showed that most of the variability in the NO densities near 100 km is due to soft x-ray flux in the 2- 7 nm. These studies (Barth et al., 1988; Siskind et al., 1990; Bailey et al., 2002) also concluded that the dissociation of $N_2$ by photoelectrons created by soft x-rays was the main mechanism of production of NO at these altitudes.

In order to effectively study the various ionization, excitation and dissociation processes induced by the photoelectrons at these low altitudes, we need the correct electron impact cross-section data together with the high-resolution solar irradiance spectrum. In this study, the high-resolution solar irradiance data is obtained from the NRLFLARE model. We have compiled a list of electron impact cross-sections from previously published experimental results (Johnson et al., 2005 and Malone et al., 2009c, 2012) and used several fitting procedures to obtain cross-sections as a function of electron energy. The compiled lists and their fitting methodology described in Sections 3.1 and 3.2 respectively, were then input into a photoelectron model. This photoelectron model uses the two-stream method introduced by Nagy and Banks (1970) and is implemented in the GLobal airglOW (GLOW) model (Solomon et al., 1988; Solomon and Abreu, 1989; Bailey et al., 2002) to calculate the photoelectron flux. From this photoelectron flux and using different electron impact ionization and excitation cross-sections data for different species, we estimate the photoelectron ionization and excitation rates by using Equation 4 (Bailey et al., 2002).

$$\int \varphi(E,z)\sigma(E)n_s(z)dE \qquad (4)$$

where $\varphi$ is the photoelectron flux, $\sigma$ is the cross-section for the process under consideration, E is electron energy, and $n_s$ is the number density of the atmospheric species undergoing the process. The NRLMSIS 2.0 (Emmert et al., 2021) model is used to obtain the neutral densities below 90 km.

The new electron impact process (excitation, ionization and dissociation) rates are then used as inputs to the Atmospheric Chemistry and Energetics (ACE) 1D model (Venkataramani et al., 2022), to quantify the global mean NO density. The ACE1D model self-consistently solves the continuity and energy equations to output the global mean neutral, ion and electron temperatures, major neutral ($N_2$, $O_2$, and O) densities, minor neutral ($N(^2D)$, $N(^4S)$, $N(^2P)$, NO, $N_2(A)$ and $O(^1D)$) densities and ion ($O^+(^4S)$, $O^+(^2D)$, $O^+(^2P)$, $N^+$, $NO^+$, $N_2^+$ and $O_2^+$) densities.

In Section 4, we discuss the results of electron impact dissociation of $N_2$ obtained using the high resolution NRLFLARE spectra and the new cross-sections and also the production of NO using the ACE1D model. Section 5 contains a discussion and comparisons of the new dissociation cross-sections and the SQ'05 cross-sections and their corresponding dissociation rates. Section 6 summarizes the results of this paper.

## 2. Photoelectron Cross-sections

As seen from Equation 4, the cross-sections for electron impact processes are required to calculate the specific photoelectron process rates. The photoelectron model can calculate the photoelectron ionization and excitation rates directly from their respective cross-sections. Experimentally, there are two main spectroscopic techniques that can be used to measure electron impact cross sections: Optical Emission Spectroscopy where the excitation cross-sections are derived from Emission Spectra (ES) (the radiation due to the transition of an electron from a higher to a lower excited state), and Electron Energy Loss Spectroscopy (EELS) where the differential cross-section is calculated from the electron energy loss spectrum of $N_2$ obtained from electron scattering experiments. The differential cross-sections are then integrated over all scattering angles to obtain the Integral Cross-sections (ICSs). To reduce uncertainty in the excitation cross-sections, we have mostly used ICS data since they are from direct measurements of inelastic collisional excitation whereas emission cross-section depends on the emission spectrum, which can be affected by cascade from higher states and branching decay channels like radiation and predissociation (Johnson et al., 2005).

Obtaining experimental DCS data and extracting ICS is not without its own uncertainties. The measurement of DCS from the $N_2$ energy loss spectrum involves the reconstruction of the spectrum in a particular energy loss range through the procedure of spectral unfolding and then a fitting analysis to derive the cross-section from the unfolded spectrum. Each of these processes have intermediary steps which adds to the uncertainty in the final result. Moreover, there are data gaps for cross-sections of all the excitation states and for a wide range of electron energy. So much so that, to date, there is no comprehensive study that provides the complete set of the relevant excitation cross-sections of $N_2$. Here we focus on compiling the most complete list of photoelectron excitation cross-sections of $N_2$ possible.

Our suggested compilation of cross-sections of $N_2$ were obtained primarily from three experimental studies conducted by Johnson et al. [2005] and Malone et al. [2009c, 2012] and is shown in Table 5. The motivation for using these newer cross-sections is to replace older DCS measurements derived from obsolete and less accurate normalization methods, insufficient angular coverage and energy resolution, and inadequate spectral unfolding techniques (Johnson et al., 2005).

Although there have been several studies that have measured the cross-section of some of the excitation states of $N_2$, to date, the Chutjian et al. [1977] and Trajmar et al. [1983] were the only published complete set of lower (with excitation energy less than 12.5 eV) and higher excited (with excitation energy 12.5 eV and greater) states of $N_2$, before the new measurements in Table 5. Trajmar et al. [1983] derived the ICS data for all low-lying states ($A^3\Sigma_u^+$, $B^3\Pi_g$, $W^3\Delta_u$, $B'\ ^3\Sigma_u^-$, a' $^1\Sigma_u$, $a^1\Pi_g$, $w^1\Delta_u$, $C^3\Pi_u$, $E'\ ^3\Sigma_g^+$, and a'' $^3\Sigma_g^+$) from DCS measured for electron energies 10 eV, 12.5, 15, 17, 20, 30, and 50 eV, covering a scattering angle of 20° to 130° in intervals of 10°, for energy loss range of 7.1-11.1 eV. The first set of ICS for $A^3\Sigma_u^+$, $B^3\Pi_g$, $W^3\Delta_u$, $B'\ ^3\Sigma_u^-$, a' $^1\Sigma_u$, $a^1\Pi_g$, and $w^1\Delta_u$, states of $N_2$ given in Table 5 were measured from DCSs of Khakoo et al. [2005], using extended energy loss range of 6.25 -11.25 eV, with an angular range of 5° to 130° with finer intervals of 5°– 10°. The ICSs were obtained for electron energies from 10 to 100 eV and extended to 200 eV for $a^1\Pi_g$ state. Khakoo et al. [2005] has provided detailed description of the spectroscopic instrumentation and the fitting procedure. Here we provide a brief summary of them that shows why these cross-sections are selected for the study of the excitation process of $N_2$. The experimental setup used cylindrical electrostatics optics and hemispherical energy selectors for the electron gun and the detector. Energy-loss spectra were obtained for the $N_2$ target gas at fixed electron impact energies and scattering angles by repetitive multichannel techniques. An improved movable source method for measuring the background signal was used (Khakoo et al., 2005) which helped to reduce the background effects by elastically scattered secondary electrons from surfaces of the apparatus and to establish a uniform transmission response for scattered electrons. The spectrometer was tuned so that the scattered electron signal followed the same relative intensity as the time-of-flight (TOF) value (which are regarded as the standard) of LeClair and Trajmar [1996]. Spectral unfolding was then applied to the spectrum to obtain the individual intensities of $A^3\Sigma_u^+$, $B^3\Pi_g$, $W^3\Delta_u$, $B'\ ^3\Sigma_u^-$, a' $^1\Sigma_u$, $a^1\Pi_g$, and $w^1\Delta_u$, and $C^3\Pi_u$ states relative to their total summed intensity. In the unfolding procedure, the obtained spectrum for each of the multi-channels, was fitted to a function of the DCS, Franck-Condon (FC) factors, instrument line function and the background signal. The FC principle governs the relative population of a given vibrational level (v′) after an electronic transition of ground state $X^1\Sigma_g^+$ (v″ = 0) to the excited state n′(v′) where n′ is the higher electronic state. The FC factors were fixed values obtained from a variety of sources (Benesch et al., 1965; Tanaka et al.,1964; Cartwright et al., 1977).

Additional energy loss spectra for elastic regions (-0.25 eV to +0.25 eV energy loss) were also obtained, together with the inelastic (6 eV to 11.5 eV energy loss) regions. The total summed DCSs were then derived by normalizing the elastic peak counts to an average of experimental scattering DCSs obtained from a number of sources (Shyn and Carignan, 1980; Trajmar et al., 1983; Nickel et al., 1988; Gote and Ehrhardt, 1995). These relative summed DCSs were placed on an absolute scale using the inelastic DCSs obtained from the TOF measurements of LeClair and Trajmar [1996] to correct for any changes in analyzer response function between the elastic and inelastic features. In this way the full DCSs of each electronic state for all vibrational levels were obtained for a particular electron energy. The errors in the DCS accounted for in Johnson et al. [2005] are: statistical and fitting errors in the individual scattering intensities, inelastic-to-elastic ratio error of the TOF results of LeClair and Trajmar [1996], the error from the available elastic scattering DCSs, propagation error due to the inelastic to elastic ratio measurements, error in the analyzer response function and an error from the flux-weighted FC factors. Some of the experimental errors were minimized by using the movable target method implemented by Khakoo et al. [2005] that reduces the background effects produced by scattering by secondary electrons from the surfaces in the apparatus. Furthermore, variation in the spectrometer response as a function of energy loss was reduced by tuning and also by using highly accurate

TOF experimental data of LeClair and Trajmar [1996] for normalization. These improvements make the DCS data of Khakoo et al. [2005] better suited for excitation studies than earlier experimental data (Trajmar et al., 1983). The ICS was then obtained just by integrating the DCS (which is a function of the electron energy and the scattering angle) over all scattering angles. The DCS was interpolated between 5° −130° and extrapolated at angles below 5° and above 130°. Interpolation between measured angles was performed using a B-spline algorithm. Due to the lack of theoretical DCS data to reliably extrapolate at experimentally inaccessible scattering angles, extrapolation was done based on general trends in DCS data. This method produces considerably less error than anticipated because the DCS is multiplied by the sine of the scattering angle ($\theta$), which at low angles between 0° and 5°, and at $\theta \rightarrow 180°$ has minimal contribution. Furthermore, the forward peak nature of the DCS also minimizes the contributions of the extrapolated regions. The uncertainty in the final ICS was obtained by performing two more integration for each transition and energy, where the extrapolations were made by keeping the DCS constant below 5° up to 0° and beyond 130° to 180°. These results were combined in quadrature and together with the average DCS uncertainty gave the final uncertainty.

It is seen that the experimental resolution, spectral unfolding technique, FC factors and the DCS elastic cross-sections used can all add to the uncertainty in the final DCS cross-sections, however, most of the errors in the calculation of the ICS comes from the measurement and estimation of the DCS themselves and not from the integration process for obtaining the ICS from DCS.

Emission based studies can also be used to derive excitation cross-sections, however, these need to account for contributions from other sources, like cascade from other excited states. An example of ES measurements is the excitation cross-sections derived by Ajello and Shemansky [1985] by analytic fitting the a$^1\Pi_g$ (3,0) emission measurement at 200 eV. Although the Ajello and Shemansky [1985] cross-sections agree well with that of Table 5 above 30 eV, below this energy, the former cross-sections are higher than these direct ICS measurements. Although Ajello and Shemansky [1985] showed their cross-sections to be within 5% of Cartwright et al., [1977] and Trajmar et al. [1983], they did not take into account the renormalization of Cartwright et al. [1977] cross-sections by Trajmar et al. [1983], but most importantly, Johnson et al. [2005] noted that the greater deviation at the lower energies could be due to larger cascade contribution and not problems with incident energy resolution at the threshold as stated by Ajello and Shemansky [1985]. Finally, excitation functions derived from the LBH emission (a$^1\Pi_g$-X$^1\Sigma_g^+$) studies of Young et al. [2010] have shown excellent agreement with the Johnson et al. [2005] a$^1\Pi_g$ cross-section. The Young et al. [2010]-derived excitation cross-sections improved the dayglow studies of the SEE (Solar Extreme ultraviolet Experiment) and GUVI (Global Ultra-Violet Imager) instruments on the TIMED (Thermosphere Ionosphere Mesosphere Energetics and Dynamics) satellite (Lean et al., 2011), over Ajello and Shemansky [1985] LBH emission-derived excitation cross-sections.

The second set of excitation cross-sections of C$^3\Pi_u$, E' $^3\Sigma_g^+$, and a'' $^3\Sigma_g^+$ states given in Table 6 were measured from 13 to 100 eV electron energies, from the DCS of Malone et al. [2009a, 2009b] with similar experimental set up and data analysis as Khakoo et al. [2005]. The energy loss range covered was also increased (10.25-12.75 eV) from that of Trajmar et al. [1983]. Moreover, significant improvement was made to the measurement of cross-section of the C$^3\Pi_u$ state from that of Johnson et al. [2005]. In Johnson et al. [2005], the spectrum covered electron energy loss only up to 11.25 eV, which meant that only the first vibrational level v'=0 was measured and the cross-sections of v'=1,2,3, and 4 levels were estimated by scaling v'=0 state using FC factors. While this procedure works for the other electronic states studied in Johnson et al. [2005], many studies (Hirabayashi and Ichimura, 1991; Zubek and King, 1994; Morozov et al., 2008) have shown the non-FC behavior in the higher vibrational levels of the $C^3\Pi_u$ state, which renders the FC factor implementation incorrect. Instead, the improved DCS was derived by applying full spectral unfolding on each of the higher vibrational levels of the $C^3\Pi_u$ state (Malone et al., 2009b). In the case of Trajmar et al. [1983], the $C^3\Pi_u$ used the traditional FC factors for measuring the transition of all vibrational levels to unfold the spectrum. This study also improves upon the vibrationally resolved excitation cross-sections measured by Zubek and King [1994], and the excitation functions derived from the emission studies by Shemansky et al. [1995]. Finally, the ICS derived for the E′ $^3\Sigma_g^+$ and a″ $^3\Sigma_g^+$ states by this study also improves upon the measurements of older Trajmar et al. [1983] with considerable differences between them.

The third important study given in Table 7 measured the ICS of the high-level (with excitation energies greater than 12.5 eV) excited states $b^1\Pi_u$, $c_3\ ^1\Pi_u$, $o_3\ ^1\Pi_u$, b' $^1\Sigma_u^+$, and $c'_4\ ^1\Sigma_u^+$ of $N_2$ in the electron energy range of 17.5 - 100 eV (Malone et al., 2012). These higher excited states are either valence states ($b^1\Pi_u$ and b' $^1\Sigma_u^+$) or form part of the Rydberg series ($c_3\ ^1\Pi_u$, $o_3\ ^1\Pi_u$, $c'_4\ ^1\Sigma_u^+$). The Rydberg state of a molecule is an excited state that follow the Rydberg formula as they converge to a ionization state of the molecule (Duncan [1965], Lofthus and Krupenie [1977]). For example, the $c'_4\ ^1\Sigma_u^+$ and $c_3\ ^1\Pi_u$ converge to the X $^2\Sigma_g^+$ state of $N_2^+$, whereas the $o_3\ ^1\Pi_u$ converge to $A^2\Pi_u$ state of $N_2^+$. The ICS were derived from the DCS data of EEL spectra from 12 to 13.82 eV for scattering angles ranging from 2° to 130° (Khakoo et al., 2008). Excitation of these states result in emissions in the Extreme Ultraviolet (EUV) range as observed in the Earth's atmosphere. Of these five states, $b^1\Pi_u$ and b' $^1\Sigma_u^+$ valence states and $c'_4\ ^1\Sigma_u^+$ (n=3) Rydberg state, via radiative decay to the ground state X $^1\Sigma_g^+$, give rise to the Birge-Hopfield I, II, and Carroll-Yoshino bands, respectively. Also, the transitions of $3p\pi_u c_3{}^1\Pi_u$, $3s\sigma_g o_3{}^1\Pi_u$ (n=3) Rydberg states give rise to the Worley-Jenkins and Worley series of Rydberg bands, respectively, although these emissions are not readily observed as they are strongly pre-dissociated (Zipf and McLaughlin, 1978). In fact, since all these higher excited states are pre-dissociated to some degree, the direct excitation approach to measure ICS data is superior and more practical than the emission based approach. There is a severe lack of ICS data for these high-lying states. Till date Chutjian et al. [1977] and later Trajmar et al. [1983] published a somewhat complete list of these cross-sections, but for only two electron energies 40 eV and 60 eV. Later Ratliff et al. [1991] added cross-sections for $b^1\Pi_u$ state for 60 eV and 100 eV electron energies. These data gaps in the high lying states is due to several factors. The spectrum of *N*2 of these states are very irregular in their ro-vibrionic energy levels and also their intensities. The Rydberg-valence ungerade states of N2 are strongly coupled with their wave-functions depended on rotational states (Liu et al., 2008; Heays et al., 2012). Stahel et al. [1983] showed that homogeneous interactions within the $^1\Sigma_u^+$ and $^1\Pi_u$ manifold states are the primary cause of most of the irregularities. Other studies (Spelsberg and Meyer, 2001; Ubachs et al., [2001], Sprengers et al., 2003) showed better agreement with experimental data when these interactions were taken into account. This is due to the fact that as these states approach the ionization continuum, their vibrionic features tend to lie very close together in energy. Therefore, very high resolution EEL spectroscopy measurements are requires to correctly unfold them. Moreover, the Rydberg-valence ungerade states of N2 are strongly coupled with their wave-functions also depended on rotational states (Liu et al., 2008; Heays et al. 2012). Hence the use of traditional invariant FC factors to unfold the spectra is incorrect, yet the DCS data of Chutjian et al. [1977] and Trajmar et al. [1983], where FC factors were used, and Ratliff et al. [1991] where diabatic Rydberg-Klein-Rees (RKR)-derived FC factors were used, are still used in determination of cross-sections today. The new Malone et al. [2012] solves these problems by unfolding the spectra in the energy-loss range 12-13.82 eV and analyzing each vibrational feature individually. A pseudo-FC factor called Relative Excitation Probability (REP), which takes into consideration the Rydberg-valence interactions, are determined for each of the vibrational levels of a particular electronic state and are then used to estimate the DCS using a procedure similar to that described in Khakoo et al. [2005] and Malone et al. [2009a, b].

The dissociation cross-sections from the excitation cross-sections are derived using experimentally-determined predissociation factors or branching ratios. The predissociation factors used to estimate the dissociation of $N_2$ are mostly calculated from various emission spectroscopic experiments by comparing the emission and excitation cross-sections (Ajello et al., 1989). For example, the emission studies conducted by Zipf and McLaughlin [1978] in the EUV range of $N_2$ are used for the emission cross-sections and pre-dissociation factors of several $^1\Pi_u$, $^1\Sigma_u^+$ valence and Rydberg states. Ajello and Shemanky [1985] estimated the pre-dissociation factor of each vibrational level of the $a^1\Pi_g$ excitation state, which is a low-lying state, from the Lyman-Birge-Hopefield (LBH) band emission at the FUV range (120-210 nm). These predissociation percentages are an average for all the vibrational states of a particular excited state of $N_2$ as, usually, different vibrational levels of a particular excited state will have different amounts of predissociation.

Earlier mass spectroscopy experiments for the electron impact ionization cross-section of $N_2$, such as Rapp et al. [1965] which are used in atmospheric models even today, deduced relative total ionization and dissociative ionization cross-section, which were then renormalized against other ionization cross-section to find their absolute values. In order to find absolute total ionization and partial cross-sections of their dissociated fragments, Straub et al. [1996] used a Time-Of-Flight (TOF) spectrometer and position sensitive detector to collect all the dissociated fragments, which otherwise escape in conventionally mass spectroscopy, due to their low kinetic energies. The cross-sections for the dissociated fragments of Straub et al. [1996] were separated into two channels: $N^+ + N_2^{2+}$ and $N^{2+}$. Improving on Straub et al. [1996], Tian and Vidal [1998] used a Focusing Time-Of-Flight (FTOF) mass spectrometer and covariance mapping mass spectroscopic technique (Frasinski et al.,1989) to separate the single, double and triple ionization of $N_2$ and their dissociated fragments. The electron impact ionization cross-sections of $N_2$ are obtained from the FTOF measurements of Tian and Vidal [1998] and are given in Table 8. The cross-sections are divided into two channels: the non-dissociative ionization ($N_2^+$) and the dissociative ionization (Table 8). The dissociation products of double ionization of $N_2^{2+}$ can be $N^{2+}$, $N^+$ and N, whereas that of triple ionization of $N_2^{3+}$ are $N^{2+}$ and $N^+$. The dissociation products of single ionization of $N_2^+$ can be $N^+$ and N. Earlier studies (Straub et al., 1996, Krishnakumar and Srivastava, 1990) were able to provide only the single ionization ($N^+ + N_2^{2+}$) and $N^{2+}$ channels. Therefore, the cross-section of the dissociative channel due to the ionization of $N_2^+$ can be estimated more accurately from Tian and Vidal [1998]. Moreover, the cross-section of N can be more accurately measured from the Tian and Vidal [1998] single-dissociative ionization channel. Therefore, the above separation makes the estimating of the dissociation rate of $N_2^+$ from its dissociation cross-section more accurate.

## 3. Fitting Electron Impact Cross-section Data

### 3.1 Methodology

The experimentally obtained cross-section data described in Section 2 are only available at specific electron energies. To use them in the photoelectron model, we fit these data into analytic expressions using different non-linear fitting procedures so as to obtain cross-sections at any electron energy. The excitation cross-sections (in units of $10^{-16}$ cm$^2$) are fitted to two analytic expressions given in Equations 5a and 5b from Tabata et al. [2006].

$$\sigma(E) = a_1 \left[\frac{E-E_{th}}{E_r}\right]^{a_2} \cdot \left(1 + \left[\frac{E-E_{th}}{a_3}\right]^{(a_2+a_4)}\right)^{-1} + a_5 \left[\frac{E-E_{th}}{E_r}\right]^{a_6.} \left(1 + \left[\frac{E-E_{th}}{a_7}\right]^{(a_6+a_8)}\right)^{-1} \quad (5a)$$

$$\sigma(E) = a_1 \left[\frac{E-E_{th}}{E_r}\right]^{a_2} \cdot \left(1 + \left[\frac{E-E_{th}}{a_3}\right]^{(a_2+a_4)}\right)^{-1} \quad (5b)$$

Where, $E_r = 1.36 \times 10^{-2} keV$ (one Rydberg), $E_{th}$ is the threshold excitation energy E is the electron energy and $a_1, a_2, \ldots a_8$ are fitting parameters.

The vibrational excitation cross-section of N$_2$ is derived using the analytic expression of Green and Barth [1965] given in Equation 6.

$$\sigma(E) = \frac{q_0 a_o}{E_{th}^2} \left(\frac{E_{th}}{E}\right)^{\Omega} \left(1 - \left(\frac{E_{th}}{E}\right)^{bb}\right)^{v} \quad (6a)$$

$$q_0 = 5.51 \times 10^{-14} cm^2 eV^2 \quad (6b)$$

Where, $a_o, \Omega, \text{bb}, \text{and } v$ are the fitting parameters.

The fitting parameters of the excitation cross-sections for the three species N$_2$, O$_2$ and O are given in Tables 1, 2 and 3, respectively.

For fitting of the ionization cross-sections, we use the analytic expression of Green and Sawada [1972] shown in Equations 7. This equation is also used to find the ionization cross-section as a result of the secondary electrons produced from the ionization process.

$$\sigma(E) = 10^{-16} a_e \Gamma_a \tan^{-1}\left(\frac{\left(\frac{E-E_{th}}{2}-T_0\right)}{\Gamma_a} + \tan^{-1}\left(\frac{T_0}{\Gamma_a}\right)\right) \quad (7a)$$

$$\text{where, } a_e = \frac{a_k}{E} \ln\left(\frac{E}{a_j}\right) \quad (7b)$$

$$\Gamma_a = \Gamma_s \frac{E}{(E+\Gamma_b)} \quad (7c)$$

$$T_0 = T_s - \frac{T_a}{(E+T_b)} \quad (7d)$$

Where, $T_s, T_a, T_b, \Gamma_s, \Gamma_b, a_k$ and $a_j$ are fitting parameters. These fitting parameters for the electron impact ionization cross-sections for the three species are given in Table 4.

**Table 1a: Fitting Parameters for $N_2$ Excitation Cross-section**

| State | $E_{th}$ (eV) | $a_0$ | $\Omega$ | bb | $v$ |
|---|---|---|---|---|---|
| Vibrational | 1.85 | 1.35 | 8 | 1 | 1.58 |

# Table 1b: Fitting Parameters for $N_2$ Excitation Cross-section

| State | $E_{th}$ (eV) | $a_1$ | $a_2$ | $a_3$ | $a_4$ | $a_5$ | $a_6$ | $a_7$ | $a_8$ |
|---|---|---|---|---|---|---|---|---|---|
| A $^3\Sigma_u^+$ | 6.17 | 9.5753 | 2.41725 | 3.48174 | 6.4202 | 1.05341 | 2.0657 | 6.74498 | 1.02078 |
| B $^3\Pi_g$ | 7.35 | 148.927 | 3.37549 | 2.62388 | 1.48676 | 0.083243 | 1.89009 | 21.5238 | 2.48385 |
| W $^3\Delta_u$ | 7.36 | 2.85419 | 3.05068 | 6.78624 | 1.31338 | 2.38554 | 20.1055 | 11.6571 | 4.35359 |
| B' $^3\Sigma_u^-$ | 8.17 | 99.6172 | 7.9575 | 0.7916 | 1.0916 | 0.5619 | 2.925 | 7.7983 | 1.262 |
| a' $^1\Sigma_u^-$ | 8.4 | 99.6141 | 8.1527 | 0.6147 | 0.8563 | 7.5378 | 4.956 | 4.9845 | 0.8058 |
| a $^1\Pi_g$ | 8.55 | 99.5923 | 8.4896 | 0.8923 | 1.7735 | 0.5584 | 1.4904 | 11.3971 | 0.8712 |
| w $^1\Delta_u$ | 8.89 | 20.3014 | 5.9437 | 5.1718 | 0.8282 | - | - | - | - |
| C $^3\Pi_u$ | 11.03 | 499.8625 | 4.9213 | 3.0122 | 2.0885 | 0.1251 | -1.4981 | 10.2382 | -0.4407 |
| E $^3\Sigma_g^+$ | 11.88 | 98.4972 | 19.0442 | 8.691 | 1.4275 | 0.0032 | -0.0748 | 8.199 | 19.9242 |
| a'' $^1\Sigma_g^+$ | 12.25 | 98.8627 | 14.6712 | 8.2492 | 0.5516 | 0.0824 | 0.7443 | 6.1166 | 3.155 |
| b $^1\Pi_u$ | 12.5 | 3.1219 | 3.0088 | 4.0583 | 0.3174 | 0.1606 | 1.1773 | 21.0129 | 0.7159 |
| c'$_4$ $^1\Sigma_u^+$ | 12.93 | 0.0728 | 1.2415 | 33.5635 | 0.4428 | 4.9739 | 3.3763 | 0.0134 | 1.9084 |
| b' $^1\Sigma_u^+$ | 12.85 | 0.1237 | 1.3607 | 22.3639 | 0.4218 | 4.9800 | 4.2974 | 0.1170 | 1.8511 |

| | | | | | | | | |
|---|---|---|---|---|---|---|---|---|
| $c_3\ ^1\Pi_u$ | 12.91 | 0.0817 | 1.1999 | 27.6399 | 0.5989 | - | - | - | - |
| $o_3\ ^1\Pi_u$ | 13.1 | 0.0334 | 0.5426 | 98.2133 | 2.4390 | 99.9950 | 3.0098 | 0.0505 | 4.0139 |
| $N_2$(15.8eV peak) | 16.4 | 0.3908 | 1.1866 | 15.6934 | 0.5019 | 0.0326 | 1.4785 | 2.4783 | 13.1506 |
| $N_2$(17.3eV peak) | 17.4 | 0.1848 | 1.0904 | 14.5947 | 0.4722 | 1.3188 | 1.6956 | 0.7298 | 8.9221 |
| $N_2$(Triplet Manifold) | 11 | 99.0223 | 6.5109 | 4.9607 | 1.4943 | 2.9187 | 3.7883 | 2.0930 | 3.6823 |
| $N_2$(VUV) | 23.7 | 0.0527 | 1.2101 | 58.9775 | 0.7044 | - | - | - | - |
| $N_2$(Ryd atoms) | 40 | 0.0237 | 0.5292 | 100.2524 | 1.0136 | - | - | - | - |

**Table 2: Fitting Parameters for $O_2$ Excitation Cross-section**

| State | $E_{th}$ (eV) | $a_1$ | $a_2$ | $a_3$ | $a_4$ | $a_5$ | $a_6$ | $a_7$ | $a_8$ |
|---|---|---|---|---|---|---|---|---|---|
| Vibrational | 0.2 | 12.9375 | 7.5566 | 10.1148 | 13.5803 | 37.1013 | 5.842 | 6.218 | 0.8618 |
| $a^1\Delta_g$ | 0.977 | 1.7862 | 2.3891 | 5.4963 | 1.7106 | - | - | - | - |
| $b^1\Sigma_g$ | 1.627 | 5.7594 | 3.8908 | 4.2622 | 1.7539 | - | - | - | - |

| State | $E_{th}$ (eV) | $a_1$ | $a_2$ | $a_3$ | $a_4$ | $a_5$ | $a_6$ | $a_7$ | $a_8$ |
|---|---|---|---|---|---|---|---|---|---|
| $A^3\Sigma_u^+$, $A'^3\Sigma_u^-$, and $c^1\Sigma_u^-$ | 4.5 | 25.4093 | 7.5201 | 5.7868 | 50.1096 | 0.9881 | 1.9706 | 6.5458 | 1.2318 |
| SRC | 6.12 | 99.58 | 7.8081 | 0.4631 | 0.7547 | 0.8460 | 0.5494 | 26.5331 | 0.7434 |
| LB | 10 | 99.5343 | 9.6625 | 4.7581 | 0.6902 | 0.0624 | 0.2952 | 31.1376 | 0.8421 |
| SB | 10.3 | 99.5481 | 9.0 | 3.2491 | 0.5718 | 0.0144 | 0.5148 | 41.2214 | 0.7297 |
| Ryds | 16 | 3.0477 | 1.5724 | 12.6191 | 0.6797 | 9.9951 | 1.1743 | 0.0018 | 4.1789 |

**Table 3: Fitting Parameters for O Excitation Cross-section**

| State | $E_{th}$ (eV) | $a_1$ | $a_2$ | $a_3$ | $a_4$ | $a_5$ | $a_6$ | $a_7$ | $a_8$ |
|---|---|---|---|---|---|---|---|---|---|
| 3s $^3S^o$ | 9.5 | 0.1186 | 1.0238 | 13.8980 | 1.1150 | 0.0274 | 0.8784 | 52.6275 | 0.5161 |
| $2p^4$ $^1D$ | 1.96 | 0.0469 | 14.0079 | 6.5948 | 0.9170 | 495.2045 | 3.7869 | 2.4143 | 0.7713 |
| 3d $^3D^o$ | 12.08 | 0.1185 | 0.8416 | 3.7233 | 0.4250 | 0.0179 | 1.7335 | 20.1638 | 1.0148 |
| 3s' $^3D^o$ | 12.53 | 0.0478 | 0.7445 | 37.1416 | 0.6746 | 0.1312 | 0.8873 | 4.2165 | 0.8589 |
| 3s'' $^3P^o$ | 14.11 | 0.0277 | 1.4506 | 32.4527 | 0.8741 | 0.1872 | 1.0028 | 2.6481 | 0.3287 |
| $2p^4$ $^1S$ | 4.18 | 0.0032 | 0.6338 | 20.0968 | 2.4291 | 0.1812 | 1.0276 | 3.8687 | 0.9775 |
| 2p $^5P^o$ | 15.65 | 0.0037 | 5.1219 | 18.5104 | 1.6939 | 0.1107 | 0.6327 | 35.0869 | 0.8162 |

| | | | | | | | | |
|---|---|---|---|---|---|---|---|---|
| 3p $^5$P | 10.73 | 0.1538 | 4.8487 | 7.2615 | 1.6631 | 0.0324 | 0.2460 | 11.9178 | 2.5099 |
| 4d $^3$D$^o$ | 12.75 | 0.0180 | 0.8349 | 34.5364 | 0.9812 | 9.8181 | 3.8669 | 1.6838 | 3.3599 |
| 5d $^3$D$^o$ | 13 | 0.0045 | 1.6012 | 27.9098 | 0.9244 | 0.0108 | 0.5718 | 12.6714 | 0.6722 |
| Rydbergs | 14 | 1.1437 | 1.0298 | 7.3643 | 0.4891 | 9.4479 | 11.5893 | 0.2458 | 10.4649 |
| 3p $^3$P | 10.98 | 0.3638 | 1.1726 | 6.6056 | 1.0698 | - | - | - | - |
| 3s $^5$S$^o$ | 9.14 | 98.4005 | 7.1143 | 4.8595 | 2.2804 | - | - | - | - |
| 4d' $^3$P$^o$ | 16.08 | 0.0006 | 7.0204 | 23.9377 | 0.5416 | - | - | - | - |

**Table 4: Fitting Parameters for Ionization Cross-section**

| Ionization | $E_{th}$ (eV) | $a_k$ | $a_j$ | $T_s$ | $T_a$ | $T_b$ | $\Gamma_s$ | $\Gamma_b$ |
|---|---|---|---|---|---|---|---|---|
| N$_2$ (Single Non-Dissociative Ionization) | 15.58 | 6.4298 | 14.2167 | -49.7604 | 830.0339 | 2182.8 | 41.3117 | -23.3569 |
| N+N$^+$ (Single Dissociative Ionization) | 22 | 1.4726 | 22.3963 | -44.8376 | -6192.0 | 398.1121 | 32.8228 | -42.3218 |
| O$_2$ (Single Non-Dissociative Ionization) | 12.10 | 1.9634 | 8.9237 | -48.6810 | -7.6249 | -35.1641 | 65.5480 | -40.3355 |
| O+O$^+$ (Single Dissociative Ionization) | 20 | 0.3287 | 0.5764 | 13.5476 | 8373.8 | 156.8474 | 47.8684 | -63.1845 |
| O (Single Ionization) | 13.6 | 4.8787 | 11.1297 | 17.4974 | 3450.8 | 121.6107 | 8.1247 | -4.1235 |

## 3.2 Results of Fitting Electron Impact Cross-section Data

Tables 5, 6, 7 and 8 list the results of fitting of various states of the major species in the form of maximum cross-section and energy of maximum cross-section. Also included in the tables are experimental study from where data was obtained to perform the fitting, along with the threshold energy of excitation or ionization ($E_{th}$) and percentage predissociation wherever applicable.

### 3.2.1 N$_2$ Electron Impact Excitation Cross-sections

The complete list of excitation states is shown in Table 5 together with their percentage predissociation. Also included in Table 5 are cross-sections for five dissociation loss channels: N$_2$ Triplet manifold, N Ryd atoms, VUV, 15.8 eV and 17.3 eV (Majeed and Stickland, 1997). These are peaks in the N$_2$ energy loss spectrum from the contribution of several excitation states that have not been resolved, yet their inclusion is important to account for their predissociation as a whole.

The photoelectrons with energies greater than 12.5 eV are able to excite the high lying states (b$^1\Pi_u$, c$_3$ $^1\Pi_u$, o$_3$ $^1\Pi_u$, b' $^1\Sigma_u^+$, and c'$_4$ $^1\Sigma_u^+$) of N$_2$, which almost entirely dissociate. Since we are dealing with photoelectrons having energies one and two orders of magnitudes greater than 12.5 eV, having sufficient knowledge about these states become extremely important. Unfortunately, the data available about these states is severely limited and plagued by uncertainty.

Table 5: Excitation States of $N_2$

| Excitation | $E_{th}$ (eV) | Energy (eV) of max cross-section | Max cross-section ($10^{-18} cm^2$) | % Pre-dissociation | References |
|---|---|---|---|---|---|
| $A^3\Sigma_u^+$ | 6.17 | 9.25 | 24.22 | 0 | Johnson et al. [2005] |
| $B^3\Pi_g$ | 7.35 | 10.25 | 31.16 | 0 | Johnson et al. [2005] |
| $W^3\Delta_u$ | 7.36 | 19.6 | 21.26 | 0 | Johnson et al. [2005] |
| $B'^3\Sigma_u^-$ | 8.16 | 17.73 | 5.98 | 0 | Johnson et al. [2005] |
| $a'^1\Sigma_u^-$ | 8.4 | 15.26 | 3.47 | 0 | Johnson et al. [2005] |
| $a^1\Pi_g$ | 8.55 | 22.77 | 22.20 | 12% above 10 eV (Ajello and Shemansky, 1985) | Johnson et al. [2005] |
| $w^1\Delta_u$ | 8.89 | 16.04 | 4.45 | 0 | Johnson et al. [2005] |
| $C^3\Pi_u$ | 11.03 | 14.52 | 26.84 | 50 (Majeed and Strickalnd, 1997) | Malone et al. [2009c] |
| $E^3\Sigma_g^+$ | 11.88 | 21.66 | 1.51 | 0 | Malone et al. [2009c] |
| a'' $^1\Sigma_g^+$ | 12.25 | 21.66 | 6.23 | 0 | Malone et al. [2009c] |
| b$^1\Pi_u$ | 12.5 | 37.53 | 18.35 | 95 (James et al., 1990) | Malone et al. [2009c] |
| c'$_4$ $^1\Sigma_u^+$ | 12.94 | 75.59 | 12.55 | 10 (Ajello et al., 1989) | Itikawa [2006] |
| b' $^1\Sigma_u^+$ | 12.85 | 56 | 14.07 | 84 (Ajello et al., 1989) | Malone et al. [2012] |
| c$_3$$^1\Pi_u$ | 12.90 | 53.27 | 10.12 | 99 (Zipf and Mc Laughlin, 1978) | Malone et al. [2012] |
| o$_3$$^1\Pi_u$ | 13.1 | 53.27 | 60.07 | 99 (Zipf and Mc Laughlin, 1978) | Malone et al. [2012] |
| N$_2$ (15.8 eV peak) | 16.4 | 43.61 | 25.18 | 100 (Majeed and Strickland, 1997) | Majeed and Strickland [1997] |
| N$_2$ (17.3 eV peak) | 17.4 | 41.48 | 10.81 | 100 (Majeed and Strickland, 1997) | Majeed and Strickland [1997] |
| N$_2$ (VUV) | 23.7 | 102.035 | 16.10 | 100 (Majeed and Strickland, 1997) | Majeed and Strickland [1997] |
| N$_2$ (N Ryd atoms) | 40 | 107.27 | 3.58 | 100 (Majeed and Strickland, 1997) | Majeed and Strickland [1997] |
| N$_2$ (N$_2$ Triplet manifold) | 11 | 16.04 | 4.45 | 100 (Majeed and Strickland, 1997) | Majeed and Strickland [1997] |

### 3.2.2 O$_2$ Electron Impact Excitation Cross-sections

Since the focus of this paper is the dissociation of N$_2$, we do not discuss the cross-sections of O$_2$ and O in detail. Photoelectron impact dissociation is more important for N$_2$ than O$_2$, which is more easily dissociated by solar photons. For completeness, we provide a brief discussion of their updated excitation and ionization cross-sections here.

Unlike $N_2$, for the $O_2$ excitation cross-sections, there are no available pre-dissociation rates that have been measured in literature. However, from Cosby, 1993, we know that the Herzberg pseudo-continuum ($c^1\Sigma_u$, $A'^3\Sigma_u^-$ and $A^3\Sigma_u^+$), Schumann– Runge Continuum (SRC) and the Rydberg states, pre-dissociate. We have used this assumption to calculate the dissociation cross-section of $O_2$ due to electron impact. The $c^1\Sigma_u^-$, $A'^3\Sigma_u^-$ and $A^3\Sigma_u^+$ states are assumed to pre-dissociate to the fragments $O(^3P) + O(^3P)$ with dissociation limit near 5.5 eV and the SRC dissociates to $O(^1D) + O(^3P)$ with dissociation limit at 7 eV (Cosby, 1993). The excitation states and their cross-sections are enumerated in Table 6.

**Table 6: Excitation States of $O_2$**

| Excitation | $E_{th}$ (eV) | Energy (eV) of max cross-section | Max cross-section ($10^{-18} cm^2$) | % Pre-dissociation | References |
|---|---|---|---|---|---|
| Vibrational | 0.2 | 10.25 | 94.34 | 0 | Noble et al. [1996] |
| $c^1\Sigma_u^-$, $A'^3\Sigma_u^-$ and $A^3\Sigma_u^+$ | 4.5 | 10.25 | 13.25 | 100 | Green et al. [2001] |
| SRC | 6.12 | 26.45 | 61.73 | 100 | Liu et al. [2018] |
| LB | 10 | 20.6 | 4.7 | 100 | Liu et al. [2018] |
| SB | 10.3 | 41.48 | 1.3 | 100 | Liu et al. [2018] |
| Ryds | 16 | 34 | 146.73 | 100 | Majeed and Strickland [1997] |
| $a^1\Delta_g$ | 0.97 | 6.75 | 10.36 | 0 | Shyn and Sweeney [1993] |
| $b^1\Sigma_g$ | 1.62 | 6.75 | 3.37 | 0 | Shyn and Sweeney [1993] |

### 3.2.3 O Electron Impact Excitation Cross-sections

There are 14 excited states of O in this model as seen in Table 7. Wherever possible we have used the most recent experimental data. However, newer O cross-section data are obtained from extensive theoretical modeling. These new data, for example, the B-spline R-matrix method implemented by Zatsarinny and Tayal [2002], and Tayal and Zatsarinny [2016], are significantly lower than the experimentally derived cross-sections. We have used these theoretical cross-sections wherever they were in good agreement with experimental but sparsely available data. Laher and Gilmore [1990] published a review of around 60 excitation states of O, since it is not possible to include all states separately, we have added a Rydberg state to include any higher excited states that were not included in the present data as done by Majeed and Strickland, 1997.

## Table 7: Excitation States of O

| Excitation | $E_{th}$ (eV) | Energy (eV) of max cross-section | Max cross-section ($10^{-18}$cm$^2$) | References |
|---|---|---|---|---|
| 3s $^3$S$^o$ | 9.5 | 26.45 | 8.63 | Johnson et al. [2003] |
| 2p$^4$ $^1$D | 1.96 | 5.25 | 44.94 | Doering [1992] |
| 3d $^3$D$^o$ | 12.08 | 32.31 | 3.51 | Johnson et al. [2003] |
| 3s' $^3$D$^o$ | 12.53 | 43.61 | 5.79 | Johnson et al. [2003] |
| 3s'' $^3$P$^o$ | 14.11 | 50.67 | 6.5 | Johnson et al. [2003] |
| 2p$^4$ $^1$S | 4.18 | 8.25 | 2.64 | Tayal and Zatsarinny [2016] |
| 2p $^5$P$^o$ | 15.65 | 41.48 | 11.04 | Tayal and Zatsarinny [2016] |
| 3p $^5$P | 10.73 | 17.73 | 2.58 | Zatsarinny and Tayal [2002] |
| 4d $^3$D$^o$ | 12.75 | 43.61 | 1.96 | Majeed and Strickland [1997] |
| 5d $^3$D$^o$ | 13 | 43.61 | 1.16 | Majeed and Strickland [1997] |
| Rydbergs | 14 | 26.45 | 32.41 | Majeed and Strickland [1997] |
| 3p $^3$P | 10.98 | 17.73 | 7.8 | Gulcicek et al. [1988] |
| 3s $^5$S$^o$ | 9.14 | 14.52 | 3.71 | Gulcicek et al. [1988] |
| 4d' $^3$P$^o$ | 16.08 | 50.67 | 2.44 | Vaughan and Doering [1988] |

**3.2.4 Electron Impact Ionization Cross-sections**

Another important mechanism of estimating the dissociation cross-section is from the ionization cross-section. In SQ'05 the dissociation from ionization was estimated assuming that the higher ionization states of $N_2$ having energies greater than approximately 25 eV (first dissociation limit of $N_2^+$), dissociate completely. However, in the updated model, there are two separate channels for ionization and dissociative ionization. This was done because there is no new experimental data for the higher states of $N_2^+$ and most of the improved ionization cross-section data separate the non-dissociative ionization and dissociative ionization channels. This analysis is still sufficient as the contribution to dissociation by electron-impact ionization is less significant than that by electron-impact excitation, as will be shown later in Section 4.2. In this study, we have used the $N_2$ and $O_2$ ionization and dissociative ionization cross-sections from Tian and Vidal [1998] where they have measured the cross-sections for the two channels of the ionization process from the threshold to 600 eV.

For the O ionization cross-section, we are using the results from the time-of-flight spectroscopic experiment performed by Thompson et al. [1994] covering an electron energy range of 14.1-2000 eV for singly-ionized O atoms. The ionization cross-sections for the three species are listed in Table 8.

Table 8: Ionization Channels of $N_2$, $O_2$ and O

| Ionization Channel | $E_{th}$ (eV) | Energy (eV) of max cross-section | Max cross-section ($10^{-18}$cm$^2$) | % Predissociation | References |
|---|---|---|---|---|---|
| $N_2$ (Single Non-Dissociative Ionization) | 15.58 | 87.82 | 198.83 | N/A | Tian and Vidal [1998] |
| N+N$^+$ (Single Dissociative Ionization) | 22 | 92.32 | 47.88 | 100% (Tian and Vidal., 1998) | Tian and Vidal [1998][51] |
| $O_2$ (Single Non-Dissociative Ionization) | 12.10 | 97.06 | 147.96 | N/A | Tian and Vidal [1998] |
| O+O$^+$ (Single Dissociative Ionization) | 20 | 112.76 | 62.16 | 100% (Tian and Vidal., 1998) | Tian and Vidal [1998] |
| O (Single Ionization) | 13.6 | 107.27 | 14.69 | N/A | Thompson et al. [1995] |

## 4. Results

### 4.1. Solar Input Irradiance

We use the NRLFLARE model (Siskind et al., 2022; Reep et al., 2022) and the X9 flare of September 6, 2017 as the input solar spectrum to analyze the dissociation of $N_2$ at the lower ionosphere from 60 km to 200 km. There are 46 wavelength bins from 0.05 to 175 nm, with bin widths from 1.8-175 nm as described in the modified EUVAC spectrum (Solomon and Qian, 2005; Woods and Rottman, 2002). The first three bins from 0.05 - 1.8 nm have been split into 12 bins (Siskind et al., 2022). The spectrum is shown in Figure 1. The figure also shows a quiet spectrum from the NRLFLARE model by essentially reducing the magnitude of the irradiance by various orders in the different bins. The first nine bins, from 0.05 - 0.8 nm of the quiet spectrum, were scaled to match the Geostationary Operational Environmental Satellite (GOES) data and the bins longer than 0.8 nm were scaled to match modified EUVAC model (Solomon and Qian, 2005) for a quiet condition, assuming a daily average F10.7 value of 85.

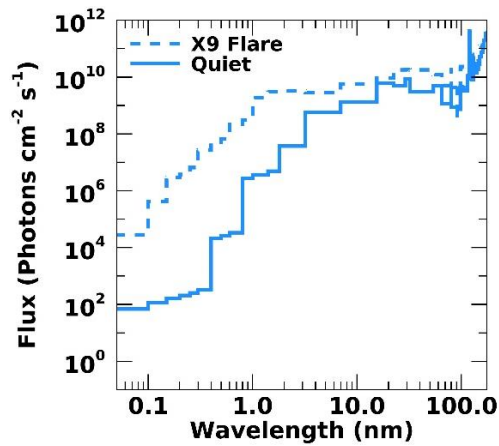

**Figure 1: Input Solar Spectra. Dashed line is the high resolution NRLFLARE Spectrum for X9 flare of September 6, 2017. Bold line is the Quiet Spectrum derived by reducing irradiance in the different bins of the NRFLARE model. The first nine bins, from 0.05-0.8 nm of the Quiet Spectrum are derived from GOES data and bins longer than 0.8 nm**

## 4.2 Updated Photoelectron Dissociation Rates

Figure 2 shows the contribution of two processes to the photoelectron dissociation of $N_2$, the electron impact inelastic collisional excitation and ionization, as estimated from their corresponding updated cross-sections. This figure highlights the importance of excitation processes in the dissociation of $N_2$. As seen from Figure 2, the dissociation due to the excitation for $N_2$ is approximately 80% of the total $N_2$ dissociation from 60 - 120 km, while the dissociation due to ionization is about 20% in the same altitude range. As we go higher up in altitude, the dissociation is mostly due to excitation (90 - 95% at 195 km) and dissociative ionization plays a minor role (5 - 10% at 195 km). Therefore, the mechanism of photoelectron dissociation is dominated by predissociation of excited $N_2$ molecule at all altitudes of interest.

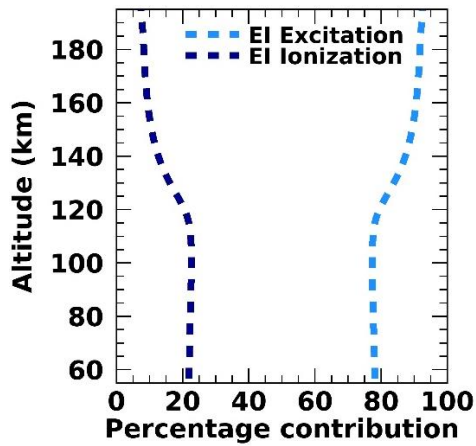

**Figure 2: Percentage contribution to dissociation due to Electron Impact (EI) excitation (dashed light blue line) and Electron Impact (EI) ionization (dashed dark blue line).**

Figure 3 shows the running integrated photoelectron spectral energy required to achieve 95% of the $N_2$ dissociation for each altitude below 200 km, for both the flare and quiet conditions. At lower altitudes near 60 km, most of the dissociation is carried out by the photoelectrons having energies of 10 keVs. At higher altitudes, i.e., above 120 km, the dissociation is mostly due to mid energy photoelectrons, about 100 eV for the quiet condition and around 500 eV for the flare. During quiet conditions, the number of photoelectrons decreases significantly. At high altitudes, since most of the dissociation is caused by low energy electrons than high energy electrons, this decrease is seen in the figure as a reduction in the energy of 95% dissociation. However, at lower altitudes, the dissociation is caused mainly by higher energy photoelectrons which are lower in number, and this percentage does not change significantly. Therefore, below 200 km, the electrons with energies greater than the 100 eV are responsible for the dissociation of $N_2$ mainly due to the predissociation of its excited states.

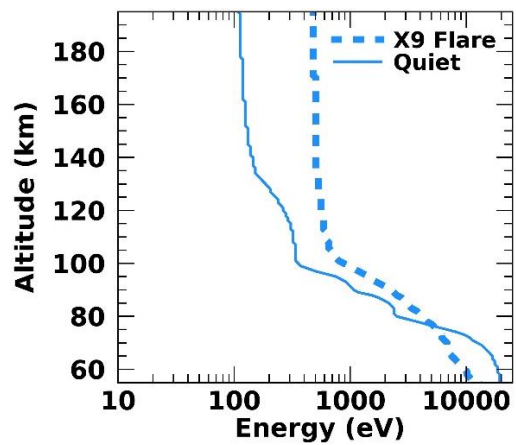

**Figure 3: Energy of 95 % dissociation of $N_2$ for Quiet (blue) and X9 flare (dashed blue line).**

Figure 4 shows the dissociation of $N_2$ in detail at two different altitudes. Figure 4a shows the photoelectron flux and the cumulative percentage dissociation of $N_2$ as a function of electron energy at altitudes of 60 km and Figure 4b shows the same at 120 km. The threshold for dissociation of $N_2$ is 10 eV, below which, the photoelectron does not have sufficient energy to dissociate the molecule. The photoelectron flux at both the altitudes are greater for low energies (near 10 eV) compared to the high energies (10 keV). This is because the photoelectrons lose their energy to atmospheric constituents as they collide, eventually cascading towards the lower energies. At the higher altitudes (120 km), the EUV radiation are absorbed by the Earth's atmosphere, where they produce photoelectrons with energies in the range of 100 eVs and lower. The soft x-rays are absorbed in the lower atmosphere near 60 km, where they produce photoelectrons in ranges of 10 keVs. During a flare event, the soft x-ray region of the solar spectrum is enhanced. For the X9 flare, at low altitudes (60 km), the difference between the low energy (around 30 eV) and high energy flux (around 10 keV) is about two orders of magnitude, whereas at the higher altitude (120 km), this difference is six orders of magnitude. Therefore, at higher altitudes, most of the dissociation is carried out by these lower energetic electrons and at lower altitudes, the dissociation is carried out by both these high and low energy electrons to an equal degree. Since the 30 eV (low energy) photoelectron flux at 60 km is about 5 orders of magnitude lower than at 120 km, but 10 keV fluxes are similar in magnitudes at both these altitudes, the effect of the high energy photoelectrons becomes visible only at these lower altitudes. This is also verified by the cumulative dissociation plots in Figure 4, which show that about 90% of the dissociation is carried out by photoelectrons with energies in 1000 eV range at 60 km and that by 100 eVs at 120 km. For the Quiet condition, the high energy photoelectron flux at 60 km almost completely vanishes (five orders of magnitude lower the flare condition).

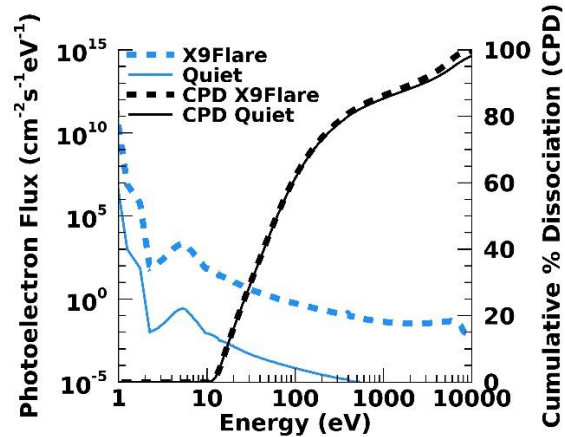

(a)

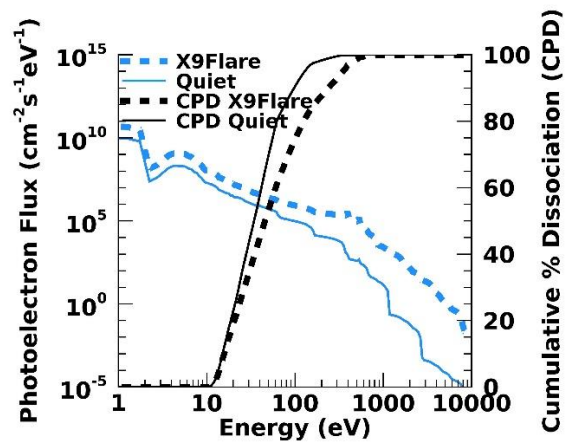

(b)

**Figure 4: Photoelectron flux (light blue) and Cumulative Percentage Dissociation (CPD) (black) for the X9 flare of September 6, 2017 (dashed) and quiet (bold) conditions at (a) 60 km and (b)120 km.**

The primary outputs of the photoelectron model for all three major species are summarized in Figure 5. Figures 5a and 5c show the photoionization ($P_i$), photoelectron ionization ($P_e$) and electron impact dissociation (EI Dissoc) rates of $N_2$ and $O_2$. The $P_e$ constitutes of both the dissociative and non-dissociative ionization and the EI Dissoc takes into account the dissociative excitation and dissociative ionization. It is important to note that since $N_2$ is a triple bond and approximately 10 eV of energy is required to dissociate it, most of the dissociation is carried out by soft x-rays and EUV radiation. However, this is not the case for $O_2$ molecule which is easily dissociated being double-bonded, therefore, there are other sources of dissociation than soft x-rays and EUV, like the Schumann-Runge band at 176 - 192.6 nm wavelengths. These other sources of dissociation for $O_2$ are not included in the photoelectron model and therefore, not shown in the figure. For a complete picture of the PE model, Figure 5e compares the photoionization and photoelectron impact ionization rates of the O atom. Figures 5b, 5d and 5f show the ratios of the photoelectron ionization and/or dissociation rate and photoionization rate for the three species. In the figures the ratio of photoelectron ionization (dissociative and non-dissociative) and photoionization rate is denoted as $P_e/P_i$ and the ratio of total dissociation and photoionization rate is denoted as EI Dissoc/$P_i$.

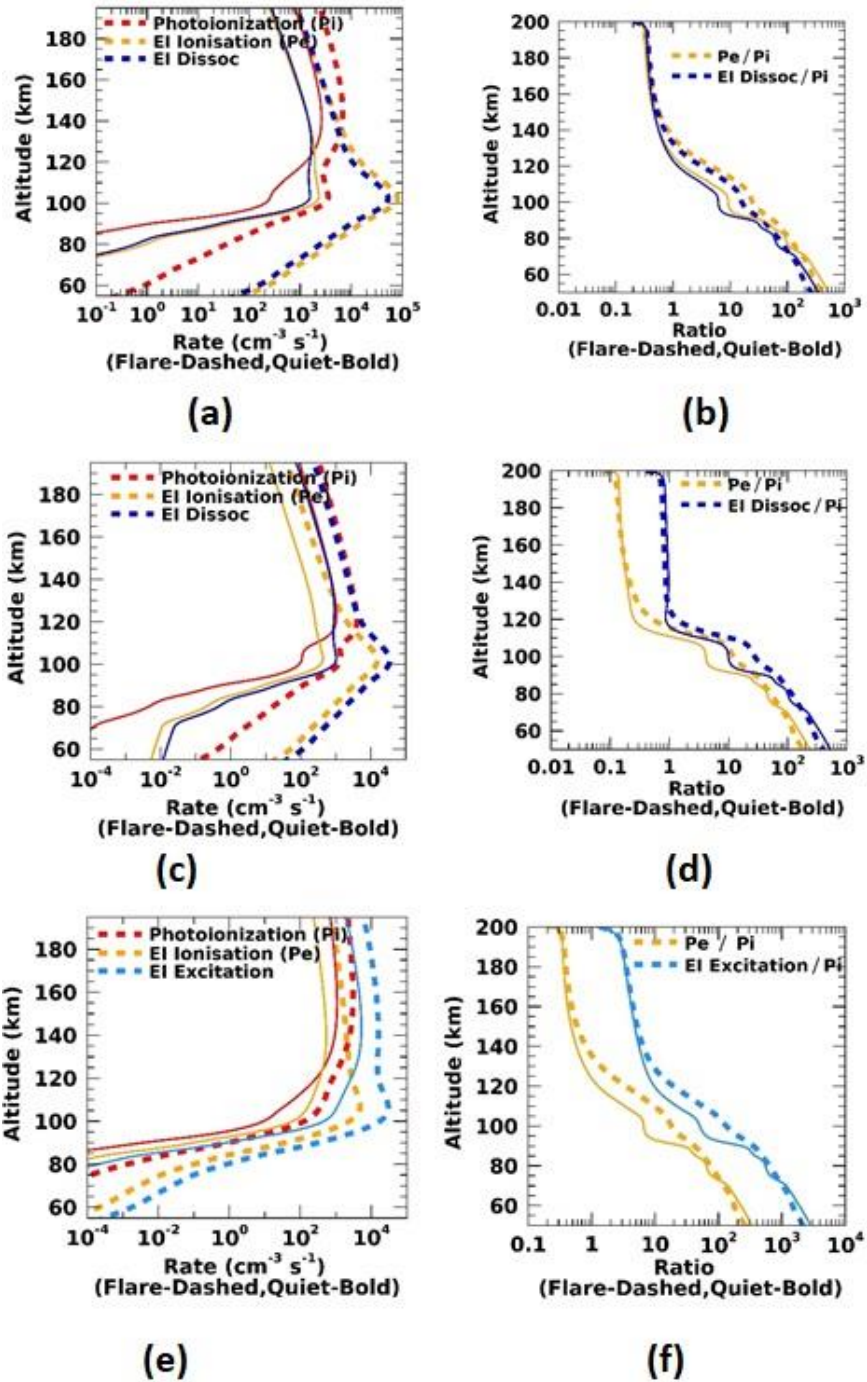

Figure 5: Photoionization, Electron Impact (EI) Excitation, EI Ionization and EI Dissociation Rates three major species (a) $N_2$, (c) $O_2$ and (e) O obtained using PE Model with the new photoelectron cross-sections. The dashed lines are from the X9 flare of September 6, 2017, while the bold lines are from the quiet condition. The ratio of each of these rates to the Photoionization ($P_i$) rates are also shown for (b) $N_2$, (d) $O_2$ and (f) O.

From Figure 5a, it can be seen that the peaks of the photoionization of $N_2$, for both the X9 flare and quiet spectra, are at about 130 km, i.e., the maximum ionization energy is deposited at this altitude. Below this there is a secondary peak at 100 km, which is very well defined for the X9 flare. This peak is therefore, due to the high energy soft x-rays photons below 1.8 nm. The various photoelectron impact rates ($P_e$ and EI Dissoc) peak at around 100 km for the flares and around 105 km for the quiet spectrum. From the dissociation curve, it can be seen that above 135 km, the $P_e$ or EI Dissoc rates are less than the $P_i$ rate. This can also be verified from Figure 5b where the Pe/Pi or EI Dissoc/Pi is less than one above 135 km, , much greater than one above 135 km., At 60 km the ratios are approximately 300. This shows that the photoelectrons that are created by the soft x-rays are the main sources of ionization of $N_2$ at the lower altitudes and are responsible for the dissociation of $N_2$ at these altitudes. Comparing the rates for the X9 flare and quiet conditions, it is seen that the photoelectron ionization and dissociation rates for $N_2$ at altitudes below 100 km varies almost two orders for magnitude and at 60 km, the quiet spectrum produces very little photoelectron ionization and dissociation. Therefore, the significance of the high energy photoelectrons is mainly seen during flares. At altitudes above 100 km the difference in photoionization and photoelectron impact ionization and dissociation rates between flares and the quiet condition are approximately an order of magnitude or lesser. Therefore, the difference in these flare and quiet rates is also a reflection of the difference in the solar input irradiance between the flare and quiet spectra. To summarize, the photoelectrons that are created by the soft x-rays are the main sources of ionization and dissociation of $N_2$ at the lower altitudes.

## 4.2 NO density from ACE1D Model

In the introduction, we discussed the mechanism for the formation of NO by the reaction of the dissociated N fragments with $O_2$ molecule. Here, we update the NO density estimation at altitudes below its peak production (106 km) (Bailey et al., 2002), by updating the electron impact cross-sections of $N_2$ and using a higher resolution solar spectrum. We used the one-dimensional ACE1D model, to estimate the NO density up to an altitude of 97 km. Since the ACE1D model gives the global average NO density, we have removed the flare component in the high-resolution solar spectrum in this part of the analysis. While estimating the NO density from Equation 3, the ACE1D model assumes constant photoelectron flux-averaged quantum yields for the three product channels of $N(^4S)$, $N(^2D)$, and $N(^2P)$ (Zipf et al., 1980; Yonker, 2013). The model uses photoelectron dissociative excitation yields of 0.5, 0.275, and 0.225 for $N(^4S)$, $N(^2D)$, and $N(^2P)$ respectively (Yonker, 2013), and the photoelectron dissociative ionization yields of 0.4 and 0.6 for $N(^4S)$ and $N(^2D)$ respectively (Venkataramani et al., 2022). The photon dissociation yields are distributed similar to the photoelectron dissociation yields.

Figure 6a shows two spectra that are used in the analysis: the modified EUVAC model, extended beyond the SQ'05 model by adding fifteen more 5 nm (105 -175 nm) bins (Woods and Rottman, 2002) and the high resolution NRLFLARE spectrum but with the flare component removed.

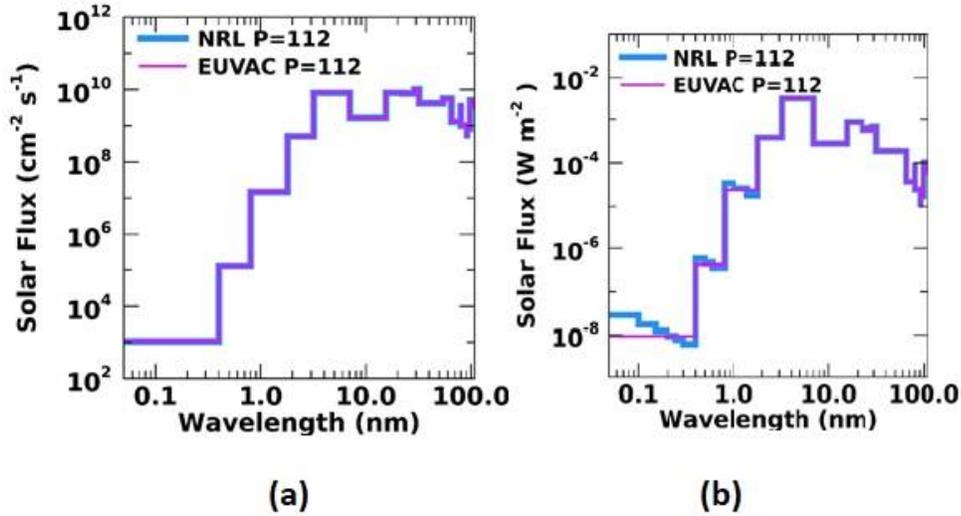

**Figure 6: Solar Spectra used in ACE1D model in units of (a) cm$^{-2}$ s$^{-1}$ and (b) W m$^{-2}$. The NRL spectra is derived from the NRLFLARE model by removing the flare component in the first twelve bins (0.05-1.8 nm). The bin widths above 1.8 nm are equal for both NRLFLARE and EUVAC model, therefore, the solar flux (in cm$^{-2}$ s$^{-1}$) in these bins are made equal to that of the EUVAC spectrum.**

The F10.7 daily and 81-day average values are the same as the September 6, 2017. The P value for this day is 112 which is low to medium. The solar flux in the twelve shortest wavelength bins from 0.05- 1.8 nm of the NRLFLARE model are reduced to the three solar flux values in the three bins, 0.05-0.4 nm, 0.4-0.8 nm, 0.8-1.8 nm, of the EUVAC model. The number of bins and bin width above 1.8 nm for both the models are the same, therefore the solar flux values in these bins for the NRLFLARE model are kept the same as the EUVAC model.

Figure 7a shows the global average NO densities as a function of altitude for three different versions of the ACE1D model: (1) The original model (Original ACE1D) with EUVAC solar spectral irradiance input and SQ'05 cross-sections, (2) Model 1, where only the photoelectron impact ionization, excitation and dissociation cross-sections and their corresponding rates have been updated but the EUVAC spectrum been used, and (3) Model 2, where both the modified NRLFLARE spectrum and the updated photoelectron impact ionization, excitation and dissociation cross-sections and rates have been used. Figure 7b shows the ratios of both these models and the original ACE1D model (Model 1/Original Model and Model 2/Original Model) as a function of altitude. As seen in Figure 7a, there are deviations in the NO densities of the three models below 120 km and the differences become more pronounced as we go lower in altitude. This can be seen clearly from the ratio plots of Figure 7b. The ratios become greater than 1 below 120 km. At the altitude of peak NO density (106 km), the NO density value of Model 1 is 1.2 times the Original model whereas the NO density of Model 2 is 1.3 times the NO densities of the Original model. Therefore, Model 1 and Model 2 are in agreement at the peak altitude. At this altitude the soft x-rays in the 1.4-7 nm bins deposit most of their energies and these bins have the same spectral irradiance as seen in Figure 6. Thus, the total contribution to the NO production from the three NRLFLARE bins (1.4-1.8 nm, 1.8-3.2 nm, 3.2-7.0 nm) and the two EUVAC bins (1.8-3.2 nm and 3.2-7 nm) at this altitude is almost equal and the deviation at this altitude is mostly due to the updating of the photoelectron cross-sections and rates. Below this altitude at 100 km, the NO density of Model 1 is 1.2 times the Original ACE1D model and the NO density of Model 2 is 1.4 times the original model. This altitude is sensitive to mostly the 0.8-1.4 nm soft x-ray bin. Since the NRLFLARE and the EUVAC spectra start to deviate from this bin as one goes towards shorter wavelengths, the difference in NO densities of Model 1 and Model 2 also increase. Below 100 km, the NO density of Model 2 is 2.3 times the original ACE1D model. This deviation largely corresponds to changes in solar soft x-rays irradiance below 1.4 nm. As we go lower in altitude (below 100 km), the deviation from the original model will grow as the higher energies short wavelength soft x-ray bins deposit their energies at these altitudes.

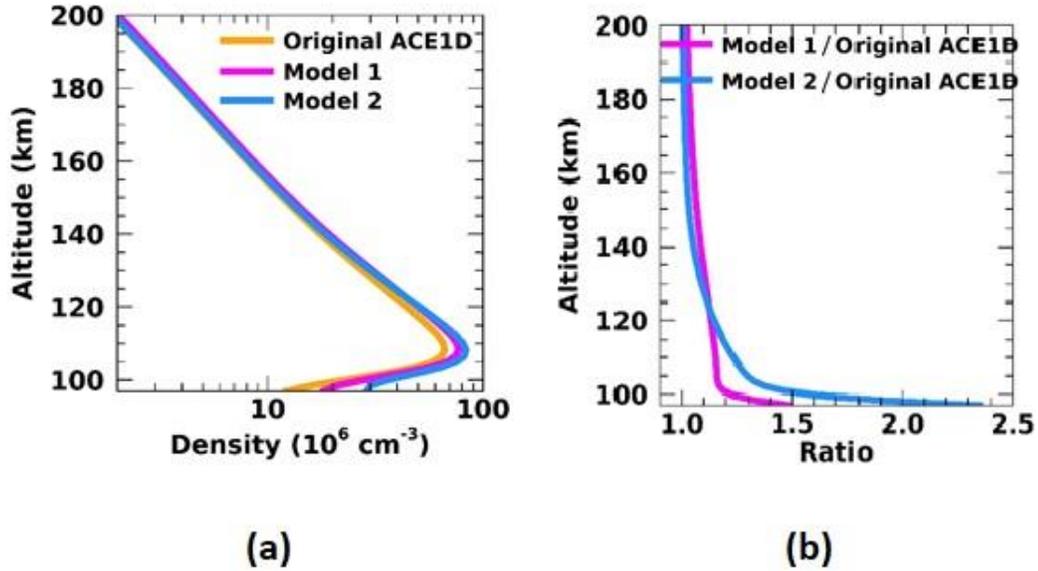

**Figure 7: (a) Global Average NO density of three versions of ACE1D model. Original ACE1D model uses the EUVAC spectrum and older $P_e/P_i$ scaling factors, Model 1 uses NRL spectrum with older $P_e/P_i$ scaling factors obtained from PE model with SQ'05 cross-sections. Model 2 uses NRL spectrum with new $P_e/P_i$ scaling factors obtained from PE model updated with new cross-sections. (b) Ratios of NO densities of Models 1 and 2 with original ACE1D model.**

5. Discussion

Figure 8 compares the electron impact dissociative excitation (Figure 8a), dissociative ionization (Figure 8b) and the total dissociation (Figure 8c) cross-sections of $N_2$ used in this study with the cross-sections from the SQ'05 model. Most of the differences are seen in the dissociative excitation cross-sections (Figure 8a). The two cross-sections show considerable difference starting from 10 eV. This is because the newer excitation cross-sections in Tables 1, 2 and 3 from which these cross-sections are derived, are lower than the SQ'05 cross-sections. The present dissociative excitation cross-sections is 0.45 times that of the SQ'05 model at about 50 eV electron energy.

The dissociative ionization cross-sections from both models, shown in Figure 8b, are in good agreement up to 100 eV, beyond which the two cross-sections start to deviate. The dissociative ionization cross-section of the present model is 0.75 times that of the SQ'05 model at 100 eV electron energy. The difference is due to the differences in the methods of estimating the present and the SQ'05 cross-sections, as explained in Section 2 yet this difference is less than the dissociative excitation cross-sections shown in Figure 8a. When using the SQ'05 model cross-sections, we assumed the excited and ionized states having threshold energies greater than 12.85 eV ($^1\Pi_u$, b'$^1\Sigma_u$, Rydberg states) and 22 eV ($D^2\Pi_g$, $C^2\Sigma_u^+$, 40 eV state), respectively. This contributes 100% to the dissociation cross-sections. In the present cross-section estimation, the dissociation from the excitation cross-sections is estimated from the experimentally determined predissociation factors, which includes lower excitation states also, which have been shown to dissociate if they have energies greater than the 10 eV dissociation energy of $N_2$.

From Figure 8c, we see that the peak dissociation cross-section is 0.65 times the SQ'05 cross-sections at about 60 eV. The difference in the total dissociation cross-section starts to decrease beyond 1000 eV. The present dissociation cross-section is on average 0.65 times the SQ'05 model cross-section. Figure 8c also shows the total dissociation cross-section obtained by Zipf and McLaughlin, [1978], which considered almost all the states considered in the current analysis. The Zipf and McLaughlin [1978] cross-sections appear closer to the SQ'05 cross-sections than the present cross-sections till 100 eV which was the maximum measured energy. However, since there is a lot of uncertainty in the electron impact cross-section data, these three results are within their measured error limits. Despite this uncertainty in the results, newer cross-sections data suggest that the excitation cross-sections and therefore, the dissociation cross-sections are lower than that obtained in these previous studies, which consequently results in lower dissociation rates of $N_2$.

Although, there is a lot of uncertainty in the electron impact cross-section data, using cross-sections data from similar experiments and adding predissociation factors help reduce uncertainties. Moreover, the present cross-sections include improved cross-sections data for the higher excited Rydberg and valence states which contribute significantly to the dissociation cross-sections of $N_2$. Although the present and the SQ'05 cross-sections, are within the error limits of the dissociation cross-sections available in literature, the updated model provides a better understanding of the dissociation process of $N_2$.

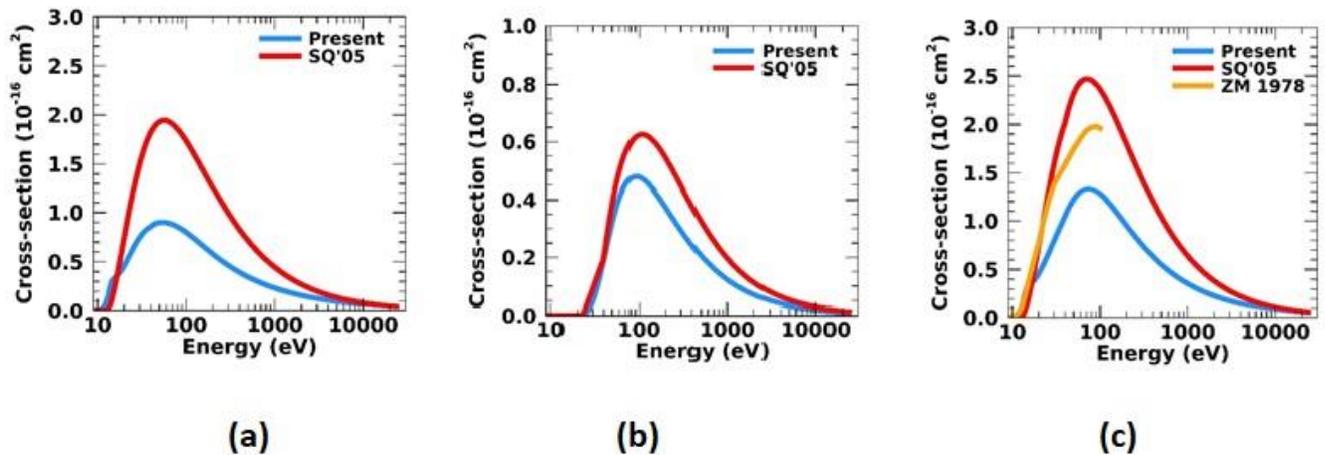

**Figure 8: Comparisons of Present (blue) and SQ'05 (red) Electron Impact (a) Dissociative Excitation (b) Dissociative Ionization and (c) Total Dissociation Cross-sections of $N_2$. ZM 1987 cross-sections (yellow) (Zipf and McLaughlin, 1978) are total dissociation cross-sections of $N_2$ for optically thin conditions**

Figure 9a shows the dissociation rates of $N_2$ as a function of altitude derived from the present and SQ'05 dissociation cross-sections shown in Figure 8c. Figure 9b shows the ratio of the two dissociation rates given in Figure 9a. The present dissociation rate is on average 0.7 times that of the SQ'05 model. Therefore, the change in the dissociation cross section as seen in Figure 8c is also reflected in the change in the dissociation rate of $N_2$ in Figure 9b.

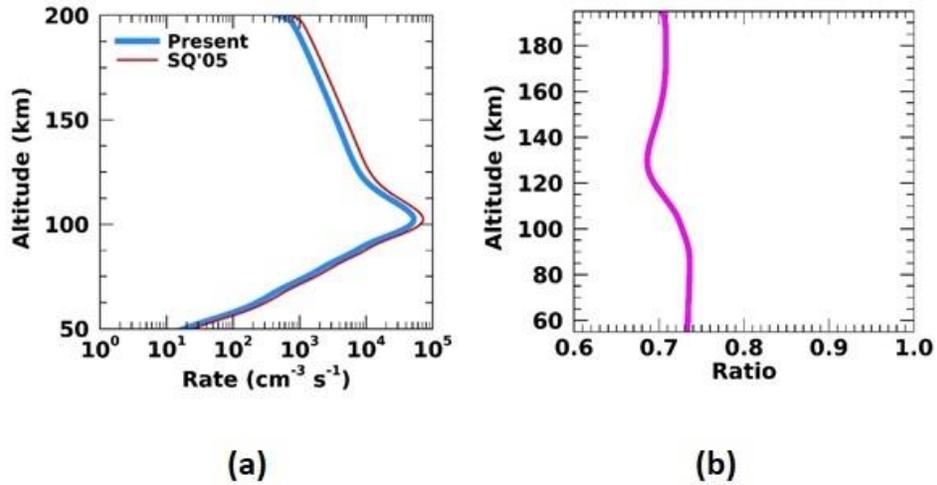

**Figure 9: (a) Comparisons of Electron Impact Dissociation Rates of $N_2$ of Present PE Model (blue line) and SQ'05 (red line) model (b) Ratio of the two models**

## 6. Summary

In this study we evaluated the process by which the solar soft x-rays in the Earth's ionosphere create highly energetic photoelectrons, which lead to the dissociation of $N_2$ molecules and the production of NO. To better quantify the $N_2$ dissociation by these photoelectrons at lower altitudes i.e., below 90 km, we revised the electron-impact excitation and ionization cross-sections used in the photoelectron model. The addition of predissociation and higher energy excited states removed some of the uncertainties in the calculation of $N_2$ dissociation rates. The dissociation cross-sections were compared to the cross-sections used in the SQ'05 model and is estimated to be 0.65 times the SQ'05 cross-sections, which is within the uncertainties that are present in the cross-sections data available in literature. The dissociation rate of $N_2$ calculated using this method is 0.7 times the dissociation rate of $N_2$ obtained using the SQ'05 model.

Using the updated photoelectron dissociation cross-sections and their corresponding dissociation rates, we estimated the global average NO densities using the one-dimensional ACE1D model. Our simulations indicate a 20% increase in the densities below the peak altitude of production than previously estimated by the model. The new NO densities can be attributed to the dissociation of $N_2$ by the high energy photoelectrons having energies in 1000 eVs- 10 keV range, which are essentially created by the soft x-rays below 2nm.

**Open Research**

The Virginia Tech photoelectron model can be found at https://github.com/baileygroup-vt under ACE-PE (Atmospheric Chemistry and Energetics- PhotoElectrons), along with the X9 flare spectrum, the quiet spectrum and the cross-sections. The ACE1D model is available at https://github.com/kvenkman/ace1d.

**Acknowledgments**